\documentclass[aps,prl,twocolumn,showpacs,preprintnumbers,amsmath,amssymb,superscriptaddress,floatfix]{revtex4-1} 

\usepackage{graphicx}
\usepackage{dcolumn}
\usepackage{bm}

\begin{document}

\title{Angular Dependence of the Superconducting Transition Temperature in Ferromagnet-Superconductor-Ferromagnet Trilayers}

\author{Jian Zhu}
\author{Ilya N. Krivorotov}
\affiliation{
Department of Physics and Astronomy, University of California, Irvine, California 92697-4575
}

\author{Klaus Halterman}
\affiliation{Research and Intelligence Department, Physics Division, Naval Air Warfare Center, China Lake, California 93555}

\author{Oriol T. Valls}
\affiliation{School of Physics and Astronomy, University of Minnesota, 
Minneapolis, Minnesota 55455}
\altaffiliation{Also at Minnesota Supercomputer Institute, University of Minnesota,
Minneapolis, Minnesota 55455}

\begin{abstract}
The superconducting transition temperature, $T_c$, of a ferromagnet (F) - superconductor (S) - ferromagnet trilayer depends on the mutual orientation of the magnetic moments of the F layers. This effect has been previously observed in F/S/F systems as a $T_c$ difference between parallel and antiparallel configurations of the F layers. Here we report measurements of $T_c$ in CuNi/Nb/CuNi trilayers as a function of the angle between the magnetic moments of the CuNi ferromagnets.
The observed angular dependence of $T_c$ is in  qualitative agreement with a F/S proximity theory that accounts for the odd triplet component of the condensate predicted to arise for non-collinear orientation of the magnetic moments of the F layers.
\end{abstract}


\maketitle

Electronic properties of a superconducting (S) film can be significantly altered by direct contact with a metallic ferromagnet (F)~\cite{bergeret2005,buzdin2005}. Penetration of the superconducting condensate into a ferromagnet results in the breaking of singlet Cooper pairs by the exchange field and leads to a decrease of the superconducting transition temperature ($T_c$). Recent discoveries of quantum interference effects in F/S heterostructures have shown that the F/S proximity effect is more interesting and complex than just the reduction of $T_c$ ~\cite{jiang1995,garifullin2002,blum2002,gu2002,tagirov1999,leksin2010,oh1997}. Examples of non-trivial F/S proximity phenomena include (i) oscillations of $T_c$ with the F layer thickness ($d_f$) in F/S multilayers~\cite{jiang1995,garifullin2002}, (ii) oscillations of the critical current in S/F/S Josephson junctions with $d_f$~\cite{blum2002} and (iii) manipulation of $T_c$ in F/S/F ~\cite{gu2002,tagirov1999} and F/N/F/S ~\cite{leksin2010,oh1997} structures (N denotes a nonmagnetic metal) by switching the magnetic state of the F layers.

The superconducting condensate of a F/S system is predicted to consist of a singlet component and a triplet component with unusual symmetry ~\cite{bergeret2001,kadigrobov2001,bergeret2003, fominov2003,volkov2003,halterman2007,halterman2008}. The triplet part of the condensate function is expected to be $s$-wave (i.e. even in the momentum variable). Therefore, by the Pauli exclusion principle, it must be odd in time. The $s$-wave character of the condensate makes it robust against impurity scattering, and thus the odd triplet pairing should survive even in dirty F/S systems. If the magnetization of the ferromagnet is not spatially uniform, the triplet part of the condensate has three non-zero components corresponding to three projections, $S_z=(0,\pm1)$ of the Cooper pair spin. The $S_z=\pm$1 components are immune to pair breaking by the exchange field, and thus can penetrate deep into the ferromagnet~\cite{bergeret2007}. Recent measurements of the critical current in Josephson junctions with noncollinear magnetic barriers support the existence of this long-range odd triplet superconductivity ~\cite{khaire2010}.

The simplest model system for studies of the effect of spatially non-uniform magnetization on superconductivity is a F/S/F trilayer with non-collinear (neither parallel nor antiparallel) orientation of the magnetic moments of the F layers~\cite{tagirov1999,fominov2003}. However, experimental tests of the F/S proximity effect in this system have been limited to the collinear geometry ~\cite{gu2002,potenza2005,moraru2006a,zhu2009,nowak2008}. In this Letter, we describe experimental and theoretical studies of $T_c$ in CuNi/ Nb/ CuNi trilayers as a function of the angle between the CuNi layer magnetizations. We compare our experimental results to theories of the F/S proximity effect that take into account the odd triplet component of the condensate, and find that the clean limit of the proximity effect theory gives good qualitative description of the data.

\begin{figure}
\includegraphics[width=0.5\textwidth]{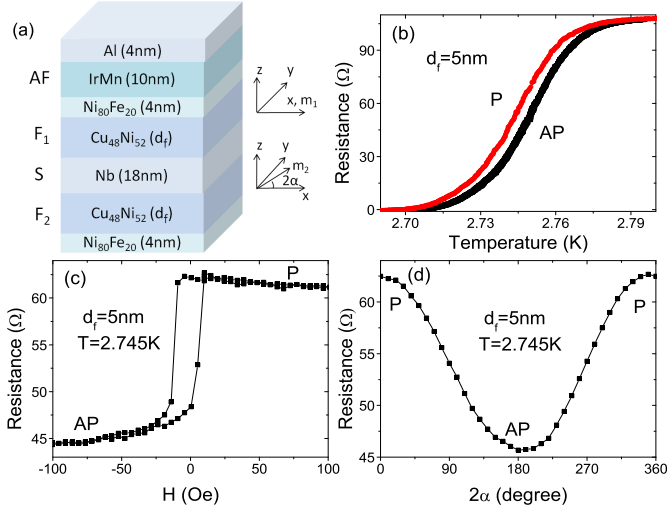}
\caption{\label{fig:fig1} (color online). (a) Schematic of the sample. Here 2$\alpha$ is the in-plane angle between the magnetic moments of the CuNi layers. (b)-(d) Experimental data for the sample with the CuNi layer thickness of 5 nm. (b) Resistance versus temperature for parallel (P) and antiparallel (AP) alignments of the CuNi layers. (c) Resistance versus in-plane magnetic field applied parallel to the pinned layer magnetization at $T$= 2.745 K. (d) Resistance versus 100 Oe in-plane field angle, $2\alpha$, measured at the same temperature.}
\end{figure}

The multilayer samples for our experiments were grown by dc magnetron sputtering in a vacuum system with a base pressure of $5.0\times10^{-9}$ Torr. We deposited a series of multilayers of the following composition: thermally oxidized Si substrate/ Ni$_{80}$Fe$_{20}$(4 nm)/ CuNi($d_f$)/ Nb($d_s=$ 18 nm)/ CuNi($d_f$)/ Ni$_{80}$Fe$_{20}$(4 nm)/ Ir$_{25}$Mn$_{75}$(10 nm)/ Al(4 nm) with $d_f$ ranging from 2 nm to 18 nm (see Fig.~\ref{fig:fig1}(a)). The ferromagnetic Ni$_{80}$Fe$_{20}$ layers placed next to the CuNi layers ensure uniform magnetization of the CuNi layers in our measurements ~\cite{gu2002}. The magnetization direction of the top F layer is pinned in the plane of the sample by exchange bias from the antiferromagnetic Ir${}_{25}$Mn${}_{75}$ layer. The CuNi alloy was grown by co-sputtering Cu and Ni targets, and the Ni concentration of 52 atomic percent was calculated from the Cu and Ni deposition rates.

We pattern the samples into 200 $\mu$m-wide Hall bars and perform four-point magnetoresistance measurements in a continuous flow ${}^4$He cryostat.  Fig.~\ref{fig:fig1}(b) shows resistance versus temperature for the parallel (P) and antiparallel (AP) configurations of the CuNi layers for the sample with 5 nm thick CuNi. The superconducting transition temperature difference between the P and AP configurations, $\Delta T_c$, is 7 mK. Fig.~\ref{fig:fig1}(c) shows the resistance as a function of magnetic field applied along the exchange bias direction in the middle of the superconducting transition at $T$= 2.745 K. This figure demonstrates that a magnetic field of $\pm$100 Oe is sufficient to reverse and saturate the magnetization of the free layer and thereby create well-defined P and AP configurations. Fig.~\ref{fig:fig1}(d) shows resistance versus the angle between the F layer magnetizations, 2$\alpha$, measured at $T$= 2.745 K. For this measurement, we stabilize the temperature in the middle of the superconducting transition to within $\pm$0.1mK, rotate the 100 Oe external magnetic field in the plane of the sample and measure the sample resistance as a function of 2$\alpha$.

\begin{figure}
\includegraphics[width=0.5\textwidth]{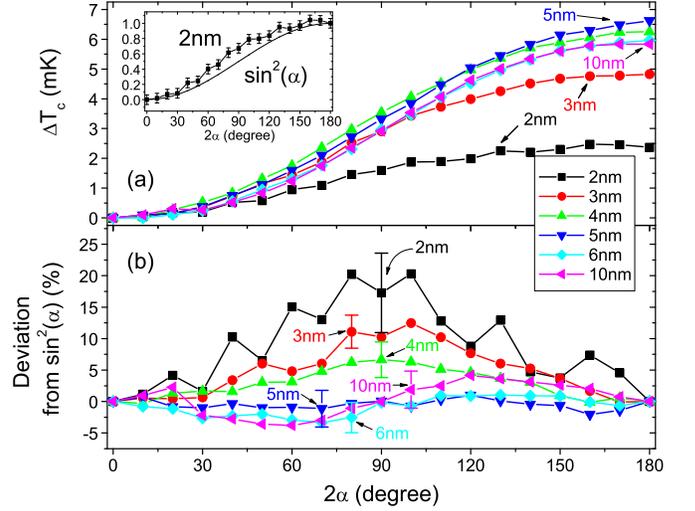}
\caption{\label{fig:fig2} (color online).
(a) Angular variation of $T_c$ defined as $\Delta T_c(d_f,2\alpha) \equiv T_c(d_f,2\alpha)-T_c(d_f,0)$ for samples with the CuNi layer thickness 
$d_f$ from 2 to 10 nm. Inset: normalized $\Delta T_c(d_f,2\alpha)$ for $d_f$ = 2 nm (squares) and $\sin^2(\alpha)$ (line). (b) Percent deviation of the normalized angular variation of $T_c$, ${\Delta T_c(d_f,2\alpha)}/{\Delta T_c(d_f,180^\circ)}$, from $\sin^2(\alpha)$.}
\end{figure}

The angular dependence of $T_c$ is calculated from the angular dependence of the resistance, $R(2\alpha)$, such as that shown in Fig.~\ref{fig:fig1}(d) \cite{note1}. In order to convert the resistance data into $T_c$, we use the slope of the $R(T)$ curve, $\left(\frac{dR}{dT}\right)$, in the middle of the superconducting transition (at $T$= 2.745 K for the sample with 5 nm thick CuNi shown in Fig.~\ref{fig:fig1}(b)). The angular variation of 
$T_c$, $\Delta T_c(d_f,2\alpha)$ = $T_c(d_f,2\alpha)-T_c(d_f,0)$, is calculated
for a given $d_f$ as $\Delta T_c(d_f,2\alpha)$ = $\left(R(2\alpha)-R(0)\right)/\left(\frac{dR}{dT}\right)$. 
Fig.~\ref{fig:fig2}(a) shows $\Delta T_c(d_f,2\alpha)$ for samples with CuNi thickness $d_f$ from 2 nm to 10 nm. This plot demonstrates that 
$\Delta T_c(d_f,2\alpha)$ significantly deviates from a simple dependence given by $\sin^2(\alpha)$. This is quantified in Fig.~\ref{fig:fig2}(b), which shows the 
percent deviation of the normalized angular variation of $T_c$, $\frac{\Delta T_c(d_f,2\alpha)}{\Delta T_c(d_f,180^\circ)}$, from $\sin^2(\alpha)$. The deviation is significant for  samples with $d_f <$ 4 nm.
The magnitude of the deviation reaches 20\% for the $d_f$= 2 nm sample.

We next discuss the comparison of our results with theory.
The transition temperature $T_c(d_f,2\alpha)$ can be calculated for our system, which is in the clean limit, using  
methods that include \cite{halterman2007,halterman2008} the odd triplet
correlations that are generated by the presence of magnets
with non-collinear magnetizations. These methods are 
based on self-consistent solution of the microscopic 
Bogoliubov-de Gennes \cite{bdg} equations. 
The matrix elements of these
equations with respect to a suitable basis containing a
sufficiently large number $N$ of elements,
can be evaluated analytically
as a function of $d_s$, $\alpha$, and the other parameters in the
problem. The eigenvalue problem for the resulting 
matrix is then numerically
solved, using \cite{halterman2008} a self consistent
iterative process. To find $T_c$,
the self-consistency equation can be linearized \cite{bvh07} near the
transition, leading
to the form $\Delta_i=\sum_q J_{iq}\Delta_q$, where the $\Delta_i$ are expansion coefficients 
of the position dependent pair potential \cite{bvh07}  in
the chosen  basis and the  $J_{iq}$ are the appropriate matrix elements with respect to the same basis, 
as obtained from the
linearization procedure. These  matrix elements can be written 
as $J_{iq} \equiv (J_{iq}^u+J_{iq}^v)/2$, where,
\begin{align}
J_{i q}^u&=
\gamma
\int d {\epsilon}_\perp
\sum_n^{N_D} \Biggl[ \tanh\biggl(\frac{ {\epsilon_n}^{u,0}}{2{T}}\biggr)
 \sum_m^N \frac{{\cal F}_{qnm} {\cal F}_{inm}}
 {{\epsilon_n}^{u,0}-{\epsilon_m}^{v,0}} \Biggr ], \\
J_{i q}^v&=
\gamma \int d  {\epsilon}_\perp
\sum_n^{N_D} \Biggl [ \tanh\biggl
( \frac{{\epsilon_n}^{v,0}}{2{T}}\biggr)
 \sum_m^N \frac{{\cal F}_{qmn} {\cal F}_{imn}}
 {{\epsilon_n}^{v,0}-{\epsilon_m}^{u,0}} \Biggr ].
\label{matfinal2}
\end{align}
Here $\gamma =\gamma_0/(4\pi D)$, with $\gamma_0$ being the usual dimensionless superconducting 
coupling, $\epsilon_n^{u(v),0}$ are unperturbed particle (hole)
energies, $D$ is the total sample thickness 
$d$ in units of the Fermi wavelength, and $N_D$ denotes that the sum is
cut off at energies beyond
the Debye frequency. 
We have 
${\cal F}_{qnm}\equiv \pi \sqrt{2 d} \sum^N_{p,r} {\cal K}_{q p r}
(d_{m p}^{\uparrow} c_{n r}^{\downarrow}+d_{m p}^{\downarrow} c_{n r}^{\uparrow})$,
${\cal K}_{q p r} \equiv (2/d)^{3/2}  \int_0^d dz f(z)\sin(k_q z) \sin (k_p z) \sin (k_r z)$, where 
$k_q = q \pi/d$,
the $z$ axis is normal to the layers (see Fig.~\ref{fig:fig1}) and
$f(z)$ is unity in the superconductor and zero elsewhere. The quantity
${\epsilon}_\perp$ denotes the kinetic energy due to
motion in the transverse direction.
The $ c_{i j}^{\sigma}$ and $d_{i j}^{\sigma}$ are the expansion coefficients
of the quasiparticle amplitudes \cite{bvh07} in terms of the basis set. 
All energies are scaled by $E_F$. 
All the sums and integrals in the above
equations can be evaluated numerically with great precision
as a function of temperature and
of the other parameters in the problem. The value
of $T_c$ is then found \cite{bvh07,lv,ad} as the highest temperature for which the largest eigenvalue of the matrix $J_{iq}$ is unity. 
This largest eigenvalue is readily found numerically. 

\begin{figure}
\includegraphics[width=0.5\textwidth]{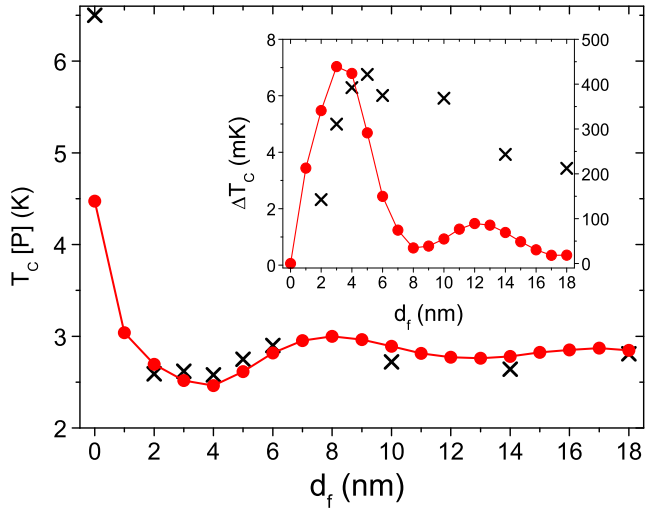}
\caption{\label{fig:fig3} (color online). Experimental data (crosses) and theoretical fittings (circles) of $T_c$ in the P state as a function of the CuNi layer thickness $d_f$. The inset is the comparison of $\Delta T_c$. Note the different scales for experimental data (left scale) and theoretical fittings (right scale).}
\end{figure}

In evaluating  $T_c(d_f,2\alpha)$ we 
have used the actual value of $d_s$ (18 nm),
and $T_{c0} =4.5$ K, where  $T_{c0}$ is the bulk transition temperature of the
material forming the superconducting layer. The only other parameters in the problem are \cite{halterman2008,bvh07}  
the superconducting coherence length $\xi_0$, a parameter $I$ characterizing the magnet strength
($I=1$ in the half metallic limit and $I=0$ for a non-magnetic
metal) and the dimensionless barrier strength \cite{hv05}, $H_B$, characterizing interfacial scattering between the S and F layers. 
The intricacies of the numerics and the computing time required
do not allow $\xi_0$, $I$, and $H_B$ to be used as free parameters in a best fit procedure. We have instead explored for best values within an experimentally reasonable region
of parameter values. Results presented here are for $\xi_0=16$ nm, 
reasonable for a Nb film, $I=0.032$ appropriate for a relatively weak magnet, and $H_B=0.75$, a  barrier of moderate strength.

The results are shown in the next two figures. In the main plot of Fig.~\ref{fig:fig3} we show the results for $T_c(d_f,0)$, the transition temperature at $\alpha=0$, versus $d_f$. The experimental results show the expected \cite{bvh07} 
oscillations with $d_f$, which are reproduced quite accurately and in
detail by the theory. In the inset, we show the 
difference $\Delta T_c(d_f,180^\circ)$ between the transition temperature at
$2\alpha=180^\circ$ and at $2\alpha=0^\circ$. This 
quantity is also plotted as a function of $d_f$.
Here, the general shape of the theoretical curve is qualitatively similar to the experiment, but the experimental overall scale of the phenomenon is much smaller than the theoretical one. We will return to this point below.

\begin{figure}
\includegraphics[width=0.5\textwidth]{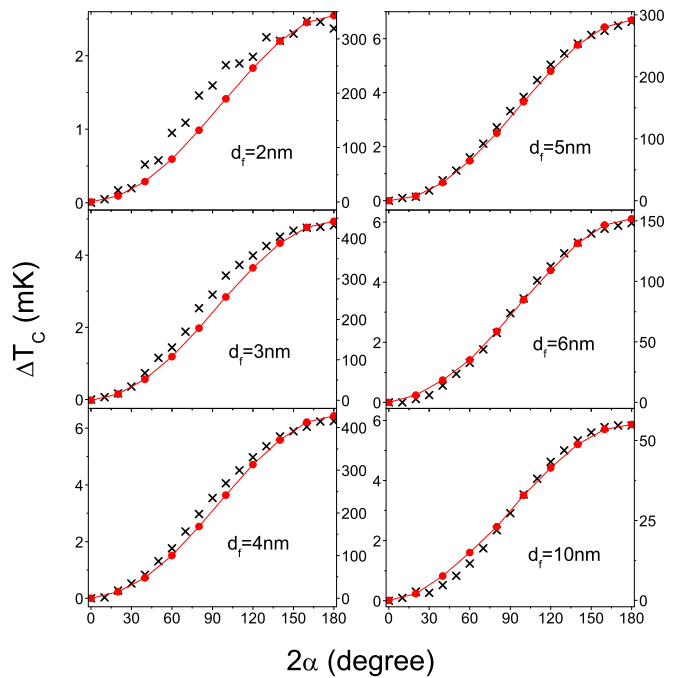}
\caption{\label{fig:fig4} (color online). Comparison between the experiment 
(crosses, left scale) and the theory (circles, right scale) of the angular dependence 
of $\Delta T_c(d_f,2\alpha)$ for samples with different CuNi layer thicknesses
$d_f$. For all $d_f$,  experimental and theoretical results have 
similar shapes despite the mismatch in absolute values. }
\end{figure}

We consider next the dependence on $2\alpha$. In the six panels of Fig.~\ref{fig:fig4} we plot 
experimental and theoretical results for $\Delta T_c(d_f,2\alpha)$ as 
a function of $2\alpha$ for six values of $d_f$ in the experimentally
relevant range. The shapes of the curves are similar to the experiment, although some deviations are present for thin ($d_f <$ 4 nm) CuNi films. 
However, the discrepancy in the magnitude found above at $2\alpha=180^\circ$ affects, unavoidably, all the other angles.

Thus, the agreement between theory and experiment is in all aspects satisfactory except for the overall scale
of the $T_c$ shifts, which is much too large in the theory. This is a puzzle,
since for apparently much more difficult properties such as the shape of
the curves in Fig.~\ref{fig:fig4} and the main plot in Fig.~\ref{fig:fig3},
theory and experiment are in good agreement. One would think,
a priori, that while the good detailed agreement in these quantities
does not preclude discrepancies elsewhere, such discrepancies
could not possibly be so large.  We have also attempted to fit the
data to the dirty limit case, using the methods of 
Ref.~\onlinecite{fominov2003}. In the dirty limit, the solution is mathematically much simpler and we are able to perform a best fit to the data using six fitting parameters. 
Despite obtaining in some cases unreasonable parameter values, the agreement with experiment for $T_c(d_f)$ (the experimental data shown
in the main plot of Fig.~\ref{fig:fig3}) 
is worse and so is the shape of the curves in the inset of 
Fig.~\ref{fig:fig3} and in Fig.~\ref{fig:fig4}, particularly at smaller values
of $d_f$. Even then, the discrepancy as to the size of the effect remains  just as large and 
in some cases even larger than in the clean limit calculation. So, the
discrepancy can not be due to the samples being closer to the dirty limit
than thought. This large discrepancy in the
magnitude of $\Delta T_c$ between theoretical predictions and experiment has been noted before ~\cite{gu2002,potenza2005,cadden2008} in other contexts.
  
It is natural to hypothesize that much larger surface scattering values, $H_B$, 
and smaller coherence length, $\xi_0$, 
leading to a diminished proximity effect, would drastically reduce 
$\Delta T_c$. This is indeed so: we find
that increasing $H_B$ and decreasing $\xi_0$ sufficiently leads to 
values of $\Delta T_c$ of the right order of magnitude, but then
the dependence of the results on $d_f$ and $\alpha$ is incorrect. A
possible explanation is that something in the sample interfaces weakens
the normal and Andreev scattering processes responsible for the
proximity effect in such a way that only the overall scale of the
amplitude of the phenomenon is changed, but we cannot speculate on what this could be.

In conclusion, we made measurements of the superconducting transition temperature $T_c$ in CuNi/Nb/CuNi trilayers as a function of the angle between the CuNi magnetic moments and the thickness of the CuNi layers. We found that $T_c$ is a monotonically increasing function of the angle between the magnetic moments of the CuNi layers. For thin ($d_f <$ 4 nm) CuNi layers, this function significantly deviates from a simple trigonometric function $\sin^2(\alpha)$. The dependence of $T_c$ on the CuNi thickness and the functional form of the angular dependence of $T_c$ are reasonably well described by the proximity effect theory in the clean limit but the overall scale of the angular variation of $T_c$ given by the theory exceed the experimentally observed values by two orders of magnitude. Our measurements of the angular dependence of $T_c$ provide new experimental constraints on theories of the proximity effect in ferromagnet/superconductor heterostructures. This work was supported by NSF Grants DMR-0748810 and ECCS-1002358. K.H. is supported in part by ONR
and by a grant of HPC resources 
as part of the DOD HPCMP.

\bibliography{tsc}

\begin{thebibliography}{29}
\bibitem [{not()}]{note1}%
  \BibitemOpen
  \href@noop {} {\bibinfo  {journal} {We assume that the pinned layer
  magnetization is fixed and that the free layer magnetization points in the
  direction of the applied field. This is a valid approximation because the
  external field (100 Oe) is much smaller than the exchange bias field ($>$1
  kOe) but much larger than the free layer anisotropy field ($<$10 Oe). We
  verified via quantitative modeling that this approximation does not have a
  significant impact on the data in Fig. [2-4].}}\BibitemShut {Stop}%
\end{thebibliography}

\end{document}